\documentclass[aps,pre,twocolumn,groupedaddress,showpacs,floatfix]{revtex4}
\usepackage{dcolumn}
\usepackage{amsmath}
\usepackage{amssymb}
\usepackage{graphicx}
\usepackage{bm}
\usepackage[T1]{fontenc}
\usepackage{color}
\usepackage{url}
\usepackage[bf]{subfigure}
\usepackage{rotating}

\begin{document}
 \title{Analysis of cluster explosive synchronization in complex networks}

 \author{Peng Ji$^{1,2}$}
 \email{pengji@pik-potsdam.de}
 \author{Thomas K.DM. Peron$^3$}
 \email{thomas.peron@usp.br}
 \author{Francisco A. Rodrigues$^4$} 
 \email{francisco@icmc.usp.br}
 \author{J\"urgen Kurths$^{1,2,5}$}
 \affiliation{$^1$Potsdam Institute for Climate Impact Research (PIK), 14473 Potsdam, Germany\\
 $^2$Department of Physics, Humboldt University, 12489 Berlin, Germany\\
 $^3$Instituto de F\'{\i}sica de S\~{a}o Carlos, Universidade de S\~{a}o Paulo, Caixa Postal 369, 13560-970, S\~{a}o Carlos,  S\~ao Paulo, Brazil\\
 $^4$Departamento de Matem\'{a}tica Aplicada e Estat\'{i}stica, Instituto de Ci\^{e}ncias Matem\'{a}ticas e de Computa\c{c}\~{a}o, Universidade de S\~{a}o Paulo, Caixa Postal 668, 13560-970 S\~{a}o Carlos,  S\~ao Paulo, Brazil\\
 $^5$Institute for Complex Systems and Mathematical Biology, University of Aberdeen, Aberdeen AB24 3UE, United Kingdom}

\begin{abstract}
Correlations between intrinsic dynamics and local topology have become a new trend in the study of synchronization in complex networks. In this paper, we investigate the influence of topology on the dynamics of networks made up of second-order Kuramoto oscillators. In particular, based on mean-field calculations, we provide a detailed investigation of cluster explosive synchronization (CES)~[Phys. Rev. Lett. 110, 218701 (2013)] in scale-free networks as a function of several topological properties. Moreover, we investigate the robustness of discontinuous transitions by including an additional quenched disorder, and we show that the phase coherence decreases with increasing strength of the quenched disorder.  These results complement the previous findings regarding CES and also fundamentally deepen the understanding of the interplay between topology and dynamics under the constraint of correlating natural frequencies and local structure.
\end{abstract}


\maketitle

 \section{Introduction}

Synchronization plays a prominent role in science, nature, social life, and engineering~\cite{Pikovsky03,strogatz2003sync,Arenas08:PR}. In recent years, much research has been devoted to investigate the effects of network topology on the emergence of synchronization~\cite{PhysRevE.87.032807, Arenas08:PR}. 
For instance, the Kuramoto oscillators undergo a second order phase transition to synchronization and the onset of synchronization is determined by the largest eigenvalue of the adjacency matrix~\cite{Restrepo05:PRE}. 

Until 2011, only continuous synchronization transitions were known to occur in networks of first-order Kuramoto oscillators~\cite{Arenas08:PR}. However,  G\'omez-Garde\~nez et al.~\cite{PhysRevLett.106.128701} reported the first observation of discontinuous phase synchronization transitions in scale-free networks, triggering further works on the subject~\cite{PhysRevLett.108.168702,PhysRevE.86.016102,
PhysRevE.86.056108,0295-5075-101-3-38002,zhang2013explosive,
su2013explosive,zou2014basin,skardal2013effects,PhysRevE.89.062811, sonnenschein2013networks}. G\'omez-Garde\~nez et al.~\cite{PhysRevLett.106.128701} considered a new kind of interplay between the connectivity pattern and the dynamics. More specifically, the authors considered the natural frequencies of the oscillators to be positively correlated with the degree distribution of the network by assigning to each node its own degree as its natural frequency, rather than drawing it from a given symmetric distribution independent of network structure, as performed in previous works~\cite{Arenas08:PR}. 

The phenomenon of explosive synchronization was proved to be an effect exclusively due to the microscopic correlation between the network topology and the intrinsic dynamics of each oscillator. 
Abrupt phase transitions were also previously observed in other dynamical processes in complex networks, such as in the context of explosive percolation in random~\cite {achlioptas2009explosive} and scale-free~\cite{radicchi2009explosive,cho2009percolation} networks. Similar to explosive synchronization, the explosive percolation has also a dynamical constraint related to the connectivity patterns, which is called the Achlioptas process~\cite{achlioptas2009explosive}. However, Costa et al.~\cite{PhysRevLett.105.255701} considered a representative model and demonstrated that the explosive percolation transition is actually a continuous, second-order phase transition with a uniquely small critical exponent of the giant cluster size.

Relevant to model several physical 
systems~\cite{PhysRevLett.109.064101,PhysRevLett.81.2229,doi:10.1137/110851584,PhysRevE.71.016215}, the second-order Kuramoto model has also been investigated under the constraint of correlation between the natural frequency and the degree distribution. Recently, we studied analytically and numerically how the inclusion of an inertia term in the second-order Kuramoto model influences the network dynamics~\cite{PhysRevLett.110.218701}. We observed a discontinuous synchronization transition as in the case of the second-order Kuramoto model in a fully connected graph with unimodal symmetric frequency distributions \cite{PhysRevLett.81.2229}. However, differently from that observed in~\cite{PhysRevLett.106.128701}, where the authors found that nodes in scale-free networks join the synchronous component abruptly at the same coupling strength, we verified that nodes perform a cascade of transitions toward the synchronous state grouped into clusters consisting of nodes with the same degree. This phenomenon is called cluster explosive synchronization (CES).

Here, we extend the previous findings presented in~\cite{PhysRevLett.110.218701}. More specifically, we analyze the parameter space for both branches in the hysteretic synchronization diagram, showing how the transition from a stable limit cycle to stable fixed points takes place as a function of the node degree and coupling strength. Furthermore, we also show that the critical coupling strength for the onset of synchronization, considering the adiabatic increasing of the coupling strength, decreases as a function of the minimum degree of the network. In addition, considering the same increase of the coupling strength, we show that the onset of synchronization decreases when the exponent of power-law degree distribution is increased. 
However, the onset of synchronization is weakly affected by the exponent of the power-law degree distribution when the coupling strength is decreased adiabatically. Finally, to address the question of how robust discontinuous transitions are against degree-frequency correlations, we include an additional quenched disorder on the natural frequencies. More precisely, we show that the phase coherence decreases, contributing to greatly increasing the irreversibility of the phase transition.


It is important to remark that different kinds of cluster-like synchronization transitions have been widely studied in the context of network theory in which patterns or sets of synchronized elements emerge~\cite{Zhou06:Chaos,pecora2013symmetries}, when ensembles of coupled oscillators are nonidentical~\cite{PhysRevE.84.036208} or under the influence of noise~\cite{PhysRevE.88.012905} or delay~\cite{d2008synchronization}. Eigenvalue decomposition can also be applied to analyze clusters of synchronized oscillators~\cite{allefeld2007eigenvalue}. Moreover, recent investigations on cluster synchronization have revealed the interplay of the symmetry in the synchronization patterns~\cite{d2008synchronization,PhysRevLett.110.174102,pecora2013symmetries}.
Here, the term  ``cluster'' is redefined and nodes with the same degree are considered to pertain to the same cluster, in contrast to the common definition of a cluster of nodes consisting of oscillators with a common phase~\cite{PhysRevE.84.036208}. Our definition is based on the dynamical behavior observed in the system composed of second-order Kuramoto oscillators whose natural frequency is correlated with the network structure~\cite{PhysRevLett.110.218701}.

This paper is organized as follows: In Sec.~\ref{sec:sec_order_Kuramoto_model}, we define the second-order Kuramoto model with correlation between the frequency and degree distributions in uncorrelated networks. Section~\ref{sec:order_parameter} is devoted to the derivation of the self-consistent equations to calculate the order parameter as a function of the coupling strength in order to determine the synchronization boundaries in Sec.~\ref{sec:parameter_space}. In Sec.~\ref{sec:analytical_results}, we present our analytical and numerical results. Our final conclusions are developed in Sec.~\ref{sec:conclusions}.

 \section{the second-order Kuramoto model}
\label{sec:sec_order_Kuramoto_model}

 \subsection{The model}

The second-order Kuramoto model consists of a population of $N$ coupled oscillators  whose dynamics are governed by phase equations of the following universal form~\cite{PhysRevLett.110.218701}:
 \begin{equation}
 \frac{d^{2}\theta_{i}}{dt^{2}}=-\alpha\frac{d\theta_{i}}{dt}+\Omega_i+\sum_{j=1}^{N}\lambda_{ij}A_{ij}\sin(\theta_{j}-\theta_{i}),
\label{Eq:second_order_kuramoto}
 \end{equation} 
 where $\theta_i$ is the phase of unit $i$ $(i=1,\ldots,N)$, $\alpha$ is the dissipation parameter, $\Omega_i$ is  the natural frequency, $\lambda_{ij}$ is the coupling strength and $A_{ij}$ is an element of the adjacency matrix $\mathbf{A}$, where $A_{ij}=1$ if the oscillators $i$ and $j$ are connected or $A_{ij}=0$, otherwise. Here, we consider a homogeneous coupling $\lambda_{ij}=\lambda$, $\forall$ $i,j$. 

In order to get analytical insights on how the topology effects the dynamics, we assume that the natural frequency $\Omega_i$ of a node $i$ is proportional to its degree according to   
\begin{equation}
\Omega_i=D(k_i-\left\langle k \right\rangle),
\label{Eq:P}
\end{equation}
where $D$ is the strength of the connection between the natural frequency and degree. 
In analogy with power grid networks modeled by the second-order Kuramoto model, the choice of $\Omega_i$ as in Eq.~(\ref{Eq:P}) assumes that 
in scale-free topologies, a high number of nodes play the role of consumers (nodes with 
$k_i < \left\langle k \right\rangle $) and nodes with high degrees play the role of power producers (nodes 
with $k_i > \left\langle k \right\rangle $). Note that the relation $\sum_j \Omega_j =0$ is satisfied, which means that the total consumed power ($\Omega_i<0$) is equivalent to the total generated power ($\Omega_i >0$). 

Substituting Eq.~(\ref{Eq:P}) in Eq.~(\ref{Eq:second_order_kuramoto}), we have~\cite{PhysRevLett.110.218701} 
\begin{equation}
 \frac{d^{2}\theta_{i}}{dt^{2}}=-\alpha\frac{d\theta_{i}}{dt}+D(k_i - \left\langle k \right\rangle)+\lambda\sum_{j=1}^{N}A_{ij}\sin(\theta_{j}-\theta_{i})
 \label{Eq:Kuramoto_with_P}.
\end{equation}
In this case, all oscillators try to rotate independently at their own natural frequencies, while the coupling $\lambda$ tends to synchronize them to a common phase. The local connection between oscillators is defined by the adjacency matrix $\textbf{A}$.

\subsection{Mean field theory}

To study the system analytically in the continuum limit, we define $\rho{(\theta,t;k)}$ as the density of oscillators with phase $\theta$ at time $t$, for a given degree $k$, which is normalized as
\begin {equation}
 \int_0^{2\pi}\rho{(\theta,t;k)} d\theta =1.
\end {equation}
In uncorrelated complex networks, the approximation $A_{ij} = k_i k_j / (N \left\langle k \right\rangle)$ is made and a randomly selected edge connects to a node with degree $k$ and phase $\theta$ at time $t$ with the probability $kP(k)\rho{(\theta,t;k)}/\langle k \rangle$, where $P(k)$ is the degree distribution and $\langle k \rangle$ is the average degree. The coupling term at the right-hand side of Eq.~(\ref{Eq:Kuramoto_with_P}) is rewritten accordingly, i.e. $\sum^{N}_{j=1} A_{ij} \sin{(\theta_j-\theta_i)} = \sum_{j=1}^N k_i k_j \sin{(\theta_j-\theta_i)} / (N \left\langle k \right \rangle)$, which in the continuum limit takes the form $k \int \int P(k') k' \rho{(\theta',t;k')} \sin{(\theta'-\theta)} d k' d\theta' / \left\langle k \right \rangle$. Thus, 
the continuum version of Eq.~(\ref{Eq:Kuramoto_with_P}) 
is given by
\begin{eqnarray}\nonumber
\frac{d^{2}\theta}{dt^{2}}  &= & -\alpha\frac{d\theta}{dt}+D(k-\left\langle k\right\rangle )\\
 & +&  \frac{\lambda k}{\left\langle k\right\rangle }\int\int k'P(k')\rho(\theta',t;k')\sin(\theta'-\theta)d\theta'dk'.
 \label{eq:motion_continuum}
\end{eqnarray}

In order to visualize the dynamics of the phases, it is natural to follow~\cite{Ichinomiya04:PRE,Restrepo05:PRE} 
and define the order parameter $r$ as $r e^{i \psi(t)} = \sum_i k_i e^{i\theta_i(t)}/\sum_i k_i$, 
where $k_i$ is the degree of the node $i$ and $\psi$ is the average phase. This order parameter is different from  $r e^{i \psi(t)} = \sum_i e^{i\theta_i(t)}/N$, which accounts for the mean-field in fully-connected graphs~\cite{kuramoto2003chemical}. 

The order parameter $r$ quantifies the phase coherence. For instance, if the initial values of $\theta$ and $\dot{\theta}$ are randomly drawn from a uniform distribution and each oscillator rotates at its intrinsic frequency, then $r\approx0$. On the other hand, if the oscillators act as a giant synchronous component, $r\approx 1$. 

In the continuum limit, the order parameter $r$ can be expressed as
\begin{equation}
 re^{i\psi}=\frac{1}{\left\langle k \right\rangle}\int\int P(k)k \rho(\theta,t;k)e^{i\theta(t)} d\theta dk.
\label{eq:r}
\end{equation}
Seeking to rewrite the continuum version in terms of the mean-field quantities $r$ and $\psi$, we multiply both sides of Eq.~(\ref{eq:r}) by $e^{-i\theta}$, take the imaginary part, and we include it in Eq.~\ref{eq:motion_continuum}, obtaining
\begin{equation}
\ddot{\theta} = -\alpha \dot{\theta} + D(k-\left\langle k \right\rangle) + k\lambda r \sin(\psi - \theta),
\label{eq:kuramoto_mean_field}
\end{equation}
which is the same equation that describes the motion of a damped driven pendulum. 

In the mean-field approach, each oscillator appears to be uncoupled from each other, and they interact with other oscillators only through the mean-field quantities $r$ and $\psi$.  The phase $\theta$ is pulled towards the mean-phase $\psi$. In the case of positive correlation between frequencies and degree, we cannot set $\psi$ as constant, since the frequency distribution is not necessarily symmetric. 


To derive sufficient conditions for synchronization, we choose the reference frame that rotates with the average phase $\psi$ of the system, i.e., we define $\phi(t)=\theta(t)-\psi(t)$. If $\dot{\phi}(t)=0$, the oscillator is synchronized with the mean field.
Defining $C(\lambda r)\equiv (\ddot{\psi}+\alpha \dot{\psi})/D$ and substituting the new variable $\phi(t)$ in 
the mean-field equation~(\ref{eq:kuramoto_mean_field}), we obtain~\cite{PhysRevLett.110.218701}
 \begin{equation}
\ddot{\phi}=-\alpha\dot{\phi}+D[k-\left\langle k \right\rangle-C(\lambda r)]- k\lambda r\sin \phi.
\label{Eq:motion_continuum_phi}
\end{equation}

\section{Order Parameter}
\label{sec:order_parameter}
The solutions of Eq.~(\ref{Eq:motion_continuum_phi}) exhibit two types of long-term behavior, depending on the size of natural frequency $D(k-\left\langle k \right\rangle-C(\lambda r))$ relative to $k\lambda r$. To obtain sufficient conditions for the existence of the synchronous solution of Eq.~(\ref{Eq:motion_continuum_phi}), we derive the self-consistent equation for the order parameter $r$, which can be written as the sum of the contribution $r_{\mathrm{lock}}$ due to the oscillators that are phase-locked to the mean-field and the contribution of non-locked drift oscillators $r_{\mathrm{drift}}$, i.e.,  $r=r_{\mathrm{lock}}+r_{\mathrm{drift}}$~\cite{Tanaka1997279}.

\subsection{Locked order parameter}

Let us assume that all locked oscillators have a degree $k$ in the range $k \in [k_1, k_2]$. These oscillators are characterized by $\dot{\phi} = \ddot{\phi} = 0$ and approach a stable fixed point defined implicitly by $\phi = \arcsin{\left(\frac{\left|D\left(k - \left\langle k \right\rangle - C(\lambda r)\right)\right|}{k\lambda r} \right)}$, which is a $k$-dependent constant phase. Correspondingly, $\rho( \phi,t;k)$ is a time-independent single-peaked distribution and 
\begin{eqnarray}
\rho(\phi;k)=\delta\left[\phi-\arcsin\left(\frac{D\left(k-\left\langle k\right\rangle -C(\lambda r)\right)}{k\lambda r}\right)\right] \nonumber \\
\mbox{ for }k\in\left[k_{1},k_{2}\right],
\end{eqnarray}
where $\delta$ is the Dirac delta function. 
Therefore, the contribution of the locked oscillators is expressed as
\begin{eqnarray}
r_{\mathrm{lock}} & = & \frac{1}{\left\langle k\right\rangle }\int_{k_{1}}^{k_{2}}\int_0^{2\pi} P(k)ke^{i\phi(t)}\nonumber\\
 &  & \delta{\left[\phi-\arcsin{\left(\frac{D(k-\langle k\rangle-C(\lambda r))}{k\lambda r}\right)}\right]}d\phi dk\nonumber,\\
\label{eq:r_locked_1}
\end{eqnarray}
whose real part yields
\begin{equation}
r_{\mathrm{lock}} =\frac{1}{\left\langle k \right\rangle}\int_{k_1}^{k_2}\left. kP(k)\sqrt{1-\left(\frac{D\left(k-\left\langle k\right\rangle -C(\lambda r)\right)}{k\lambda r}\right)^2} \right. dk . 
\label{Eq:int_r_lock_real}
\end{equation} 
We consider first a scale-free network with a degree distribution given by $P(k)=A(\gamma)k^{-\gamma}$, where $A(\gamma)$ 
is the normalization factor and $\gamma=3$. Substituting the degree distribution $P(k)$ and applying the variable transformation
 $x(k)=D(k-\left\langle k \right\rangle - C(\lambda r))/\lambda k r$, we obtain the following implicit equation for the contribution of the locked oscillators

\begin{eqnarray}\nonumber
r_{\mathrm{lock}}  &=&  \frac{A(\gamma)}{2D\left\langle k\right\rangle }\left[\left(x(k_{2})\sqrt{1-x^{2}(k_{2})}\right)+\arcsin x(k_{2})\right. \\
  &-&  \left.\left(x(k_{1})\sqrt{1-x^{2}(k_{1})}+\arcsin x(k_{2})\right)\right].
 \label{Eq:r_lock_I}
\end{eqnarray}
\subsection{Drift order parameter}
We analyze the drifting oscillators for $k\in k_{\mathrm{drift}} \equiv \left[k_{\min},k_1\right]\cup \left[k_2,k_{\max}\right]$, where $k_{\min}$ denotes the minimal degree and $k_{\max}$ is the maximal degree. The phase of the drifting oscillators rotates with period $T$ in the stationary state, 
so that their density
$\rho( \phi,t;k)$ satisfies $\rho \sim |\dot{\phi}|^{-1}$~\cite{Tanaka1997279}. As
$\oint \rho(\phi;k) d\phi =\int^{T}_0 \rho(\phi;k) \dot\phi dt =1 $, this implies
$\rho(\phi;k)=T^{-1}|\dot\phi|^{-1}=\frac{\Omega}{2\pi}|\dot\phi|^{-1}$, where $\Omega$ is the oscillating frequency of the running periodic solution of $\phi$~\cite{Tanaka1997279}.
After substituting $\rho(\phi;k)$ into Eq.~(\ref{eq:r}),
 we get
\begin{equation}
r_{\mathrm{drift}}=\frac{1}{2\pi \left\langle k \right\rangle}\int_{k \in k_{\mathrm{drift}}}\int^T_0 kP(k) \Omega |\dot\phi|^{-1} e^{i\phi(t)} \dot{\phi}dt dk.
\label{eq:r_drift_1}
\end{equation}
Without loss of generality, we assume that $\dot{\phi}<0$ for $k \in \left[ k_{\min},k_1\right]$ and $\dot{\phi}>0$ for $k \in \left[k_2,k_{\max}\right]$. Thus the real part of equation~(\ref{eq:r_drift_1}) becomes
\begin{equation}
r_{\mathrm{drift}}= \frac{1}{2\pi \left\langle k \right\rangle}\left(-\int_{k_{\min}}^{k_1} + \int_{k_{2}}^{k_{\max}}\right)\int^T_0 kP(k) \Omega  \cos{(\phi)}dt dk.
\label{eq:r_drist_min_plus}
\end{equation}
A perturbation approximation of the self-consistent equations enables us to treat Eq.~(\ref{eq:r_drist_min_plus}) analytically. After performing some manipulations motivated by~\cite{Tanaka1997279}, we get
\begin{equation}
r_{\mathrm{drift}} = \left(-\int_{k_{\min}}^{k_1} + \int_{k_2}^{k_{\max}} \right)\frac{-rk^2\lambda \alpha^4 P(k)}{D^3\left[k - \left\langle k \right\rangle -C(\lambda r)\right]^3\left\langle k\right\rangle} dk
\label{Eq:int_r_drift_real}
\end{equation}
Thus, the self-consistent equation for the order parameter $r$ is obtained by summing the contribution of locked and drifting oscillator as 
\begin{equation}
r=r_{\mathrm{lock}}+r_{\mathrm{drift}}, 
\label{eq:r_rlock_rdrift}
\end{equation}
which are obtained from Eqs.~(\ref{Eq:int_r_lock_real}) and~(\ref{Eq:int_r_drift_real}), respectively.

\subsection{Determining $C(\lambda r)$}

The summation of Eq.~(\ref{Eq:int_r_lock_real}) and Eq.~(\ref{Eq:int_r_drift_real}) gives us the analytical solution for the order parameter $r$. However, there is a quantity to be determined, namely the term $C(\lambda r )$. Considering the sum of Eqs.~(\ref{eq:r_locked_1}) and~(\ref{eq:r_drift_1}) and taking its imaginary part, we get

\begin{eqnarray}
0 & = & \frac{1}{\left\langle k\right\rangle }\int_{k_{1}}^{k_{2}}kP(k)\frac{D(k-\langle k\rangle-C(\lambda r))}{k\lambda r}dk\label{eq:imaginary_part}\nonumber\\
 & +&\frac{1}{2\pi\left\langle k\right\rangle }\left(-\int_{k_{\min}}^{k_{1}}+\int_{k_{2}}^{k_{\max}}\right)\int_{0}^{T}kP(k)\Omega|\dot{\phi}|^{-1}\sin\phi dtdk.\nonumber\\
\end{eqnarray}
Following a similar procedure to approximate $\int_0^T \cos\phi(t) dt$ in Eq.~(\ref{eq:r_drist_min_plus})~\cite{Tanaka1997279} for the integral $\int_0^T \sin\phi(t) dt$ in Eq.~(\ref{eq:imaginary_part}), we obtain

\begin{eqnarray}
0 & = & \frac{1}{\left\langle k\right\rangle }\int_{k_{1}}^{k_{2}}kP(k)\frac{D(k-\langle k\rangle-C(\lambda r))}{k\lambda r}dk\nonumber\\
 &  & +\frac{1}{2\left\langle k\right\rangle }\left(\int_{k_{\min}}^{k_{1}}+\int_{k_{2}}^{k_{\max}}\right)\frac{rk^{2}\lambda\alpha^{2}P(k)}{D^{2}\left[k-\left\langle k\right\rangle -C(\lambda r)\right]^{2}}dk\nonumber\\
 \label{eq:equation_for_C}
\end{eqnarray}
Therefore, through Eq.~(\ref{eq:equation_for_C}) we yield the evolution of $C(\lambda r)$ as a function of the coupling $\lambda$, and then, together with Eqs.~(\ref{Eq:int_r_lock_real}) and (\ref{Eq:int_r_drift_real}), we have the full recipe to calculate the order parameter $r$.\\

\section{Parameter space and synchronized boundaries}
\label{sec:parameter_space}
It is known that systems governed by the equations of motion
 given by Eq.~(\ref{Eq:motion_continuum_phi}) present a hysteresis as $\lambda$ is varied~\cite{strogatz1994nonlinear,tanaka1997first,Tanaka1997279}. 
Therefore we consider two distinct cases: 
(i) Increase of the coupling strength $\lambda$.
In this case, the system starts without synchrony ($r\approx 0$) and, as $\lambda$ is increased, approaches the synchronous state ($r\approx 1$). 
(ii) Decrease of the coupling strength $\lambda$.
Now the system starts at the synchronous state ($r\approx 1$) and, as the $\lambda$ is decreased, 
more and more oscillators lose synchrony, falling into the drift state. 

Next, we study the following problem: why do phase transitions occur for a continuously varying coupling strength? 
 We illustrate the phase transitions using the parameter space of the pendulum. 
For convenience, we non-dimensionalize Eq.~(\ref{Eq:motion_continuum_phi}) by $\tau = \sqrt{k\lambda r}t$~\cite{Strogatz2001}, and set $\beta \equiv \alpha / \sqrt{k\lambda r}$ and 
$I \equiv D(k - \left\langle k \right\rangle - C(\lambda r))/(k\lambda r)$, 
yielding the dimensionless version:
\begin{equation}
\frac{d^2\phi}{d^2\tau} + \beta \frac{d\phi}{d\tau} + \sin \phi = I.
\label{Eq:kuramoto_mean_field_new_time_scale}
\end{equation}
The variable  
$\beta$ is the damping strength and $I$ corresponds to a constant torque 
(cf. a damped driven pendulum). 
The bifurcation diagram in the $\beta-I$ parameter space of Eq.~(\ref{Eq:kuramoto_mean_field_new_time_scale}) has three types of bifurcations~\cite{strogatz1994nonlinear}: homoclinic and infinite-period bifurcations periodic orbits, and a saddle-node bifurcation of fixed points. An analytical approximation for the homoclinic bifurcation curve for small $\beta$ was derived using Melnikov's method~\cite{guckenheimer1983nonlinear,strogatz1994nonlinear} and the curve is tangent to the line $I=4\beta/\pi$. 

The parameter space is divided into three different areas corresponding to the stable fixed point, the stable limit cycle and bistability. 
When $I > 1$ or $D(k-\left\langle k \right \rangle) > k \lambda r$ in Eq. (7), in the stable limit cycle area, there is no stable fixed point  and the oscillators evolve to the stable limit cycle, regardless of the initial values of $\theta$ and $\dot{\theta}$. Therefore, in this case, the oscillators are drifting and contribute to $r_{\mathrm{drift}}$.
When $I < 1$ and $I$ is below the homoclinic bifurcation curve, only stable fixed points exist and the oscillators converge to the stable fixed points and contribute to $r_{\mathrm{lock}}$, regardless of the initial values. Otherwise, depending on the situation of the decreasing or increasing coupling strength, the oscillators within the bistable area converge to the stable fixed point (contributing to $r_{\mathrm{lock}}$) or the stable limit cycle (contributing to $r_{\mathrm{drift}}$), respectively.

Our change of time-scale allows us to employ Melnikov's analysis to determine
the range of integration $\left[k_1, k_2 \right]$ in the calculation of $r = r_{\mathrm{lock}} + r_{\mathrm{drift}}$.

\begin{figure}
\includegraphics[width=1.0\linewidth]{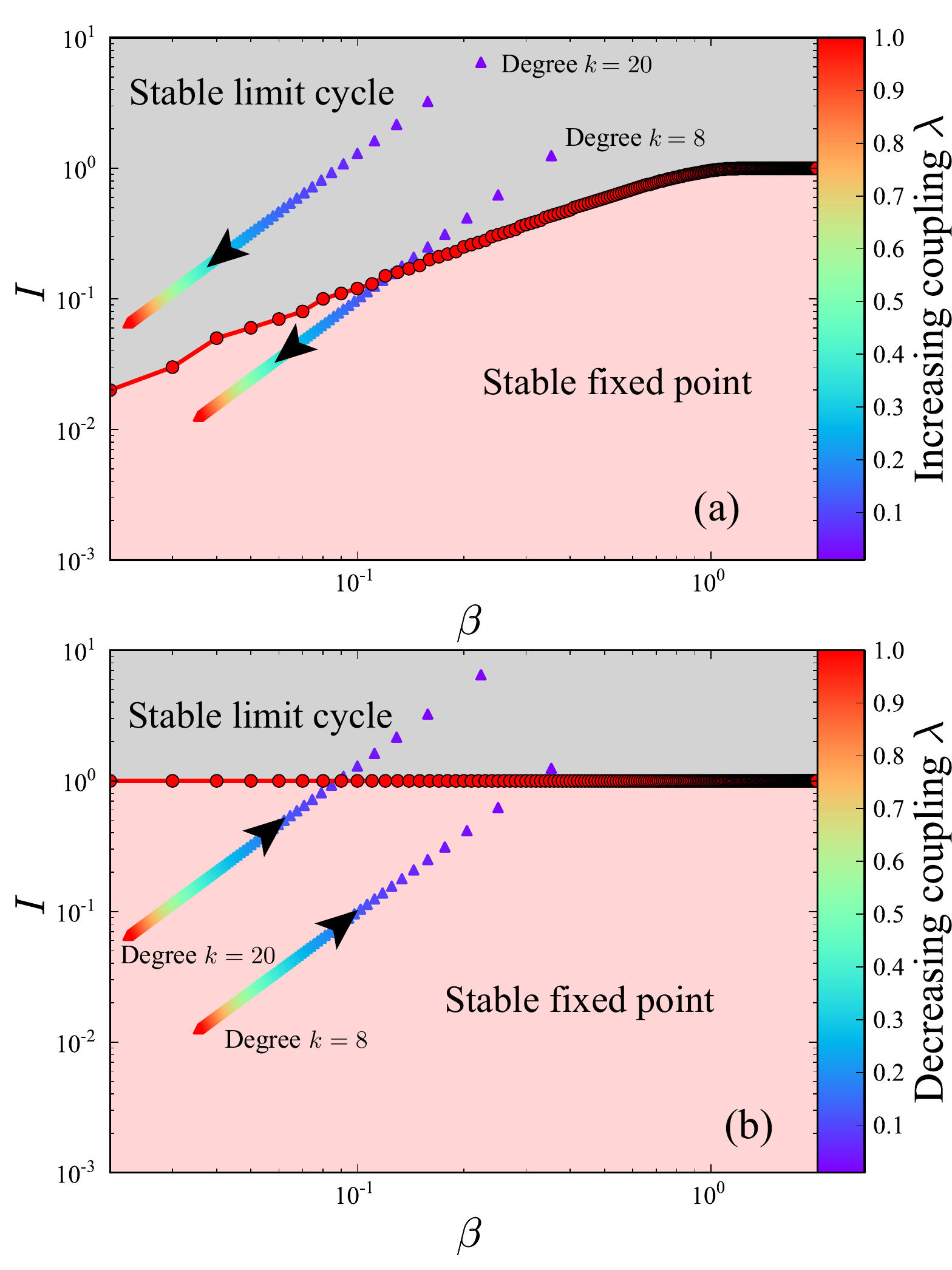}
 \caption{(Color online). Parameter space of the pendulum [Eq.~(\ref{Eq:kuramoto_mean_field_new_time_scale})]: (a) for increasing coupling strength and (b) decreasing coupling strength. 
 The red (dark gray) area indicates the existence of a stable fixed point, whereas the gray area indicates the parameter combinations that give rise to a stable limit cycle. The dots in the parameter space represent oscillators with degree $k=8$ and degree $k=20$, which start with incoherence in (a) [coherence in (b)], and approach synchronous states (incoherence), for increasing (decreasing) coupling strength $\lambda$ with $\alpha=0.1$, $D=0.1$, $\left \langle k \right\rangle=10$, $C=-3$, $N=3000$, and $P(k)\backsim k^{-\gamma}$, where $\gamma=3$.} 
 \label{fig:parameter_space_increasing_decreasing}
\end{figure} 
\subsection{Increasing coupling: Synchronized boundary}
When the coupling strength $\lambda$ is increased from $\lambda_0$, the synchronous state emerges after a threshold $\lambda_c^I$ has been crossed. Here we derive self-consistent equations that allow us to compute $\lambda_c^I$. 

The stable fixed point and the stable limit cycle coexist in the bistable area. Whether the oscillator will converge to the fixed point or rotate periodically depends crucially on the initial values of $\theta$ and $\dot{\theta}$ for given parameter values of $\beta$ and $I$. As the coupling strength increases, the bistable area vanishes and we only get the stable limit cycle in this region. The stability diagram for the increasing case is shown in Fig.~\ref{fig:parameter_space_increasing_decreasing}(a). Therefore, as we can see from this figure, for $I>1$, Eq.~(\ref{Eq:kuramoto_mean_field_new_time_scale}) has only one stable limit cycle solution. If $4\beta/\pi \leq I \leq 1$, the system is no longer bistable and only the limit cycle solution exists. If the coupling strength is increased further, 
the synchronized state can only exist for $I\leq 4\beta / \pi$, where 
Eq.~(\ref{Eq:kuramoto_mean_field_new_time_scale}) has a stable fixed point solution $\sin{(\phi)}=I$. Solving the inequalities 
\begin{equation}
\frac{\left|D\left(k - \left\langle k \right\rangle - C(\lambda r)\right)\right|}{k\lambda r}\leq 1,
\label{Eq:sin_phi_increase}
\end{equation} 
and 
\begin{equation}
\frac{\left|D\left(k - \left\langle k \right\rangle - C(\lambda r)\right)\right|}{k\lambda r}\leq \frac{4\alpha}{\pi \sqrt{k\lambda r}},
\label{Eq:Melkonivos}
\end{equation}
we get the following range of $k^I$ for the phase-locked oscillators
\begin{eqnarray}
k^I \in \left[k_1^I,k_2^I\right]=
\begin{cases}
\left[\frac{\left\langle k \right\rangle +C(\lambda r)}{1+\lambda r},\frac{\left\langle k \right\rangle +C(\lambda r)}{1-\lambda r}\right] &  \mathrm{if} \, \lambda r< b,
\\
\\
\left[\frac{\left\langle k \right\rangle +C(\lambda r)}{1+\lambda r},K_2^I\right] &  \mathrm{if} \, b<\lambda r<1,\\
\\
\left[K_1^I,K_2^I\right] & \mbox{otherwise},
\end{cases}
\label{Eq:synchronized_boundary_increasing}
\end{eqnarray}
where $b=\frac{16\alpha^2}{\pi^2[\left\langle k \right\rangle +C(\lambda r)]+16\alpha^2}$ and 
\begin{eqnarray*}\nonumber
\left[K_1^I,K_2^I\right] \equiv  \left[\frac{B-\sqrt{B^{2}-4D^{4}\left(\left\langle k\right\rangle +C(\lambda r)\right)^{2}}}{2D^{2}}\right.,\\
  \left.\frac{B+\sqrt{B^{2}-4D^{4}\left(\left\langle k\right\rangle +C(\lambda r)\right)^{2}}}{2D^{2}}\right],
 \label{Eq:k_second_range}
\end{eqnarray*}
where
\begin{equation}
B = 2D^2(\left\langle k \right\rangle + C(\lambda r)) + \frac{16\alpha^2 \lambda r}{\pi^2}.
\label{Eq:constant_B}
\end{equation}
Since $\lambda r$ is present in all equations, we define a new variable $y = \lambda r$ and analyze the self-consistent equations computing $r= y/\lambda$.

In order to visualize the dynamics and deepen the understanding of phase transitions, we sketch in Fig.~\ref{fig:parameter_space_increasing_decreasing}(a) the phase trajectories of two randomly selected oscillators with degree $k=8$ and $20$. 
When the coupling strength is close to $0$, the oscillators are in the stable limit cycle area and each node oscillates with their own natural frequency. One can see that the critical coupling for the onset of synchronization of the oscillator with degree $k=8$ is lower and thus the small degree oscillator converges to the fixed point at lower coupling strength. 

\subsection{Decreasing coupling: Synchronized boundary}
With a decreasing coupling strength $\lambda$, the oscillators 
start from the phase-locked synchronous state and reach the asynchronous state at a critical coupling 
$\lambda_c^{D}$. 
In order to calculate this threshold, we again investigate the range of degree $k^D$ of the phase-locked oscillators.
Imposing the phase locked solution 
in Eq.~(\ref{Eq:motion_continuum_phi}), we obtain $\sin\phi = \frac{\left|D\left(k - \left\langle k \right\rangle - C(\lambda r)\right)\right|}{k\lambda r}\leq 1$ 
and find that the locked oscillators are the nodes with degree $k$ in the following range as a function of $\lambda r$:
\begin{equation}
    k^D \in [k_1^D, k_2^D] \equiv \left[\frac{\left\langle k \right\rangle + C(\lambda r)}{1 + \frac{\lambda r}{D}} , \frac{\left\langle k \right\rangle + C(\lambda r)}{1 - \frac{\lambda r}{D}} \right],
\label{Eq:range_k_decrease}
\end{equation}
 when $\lambda r < D$, or $ k_1^D = \frac{\left\langle k \right \rangle + C(\lambda r)}{1+\frac{\lambda r}{D}}$ and $k_2^D \rightarrow k_{\max}$ otherwise. 
This allows us to calculate  $r^D$ and $\lambda^D_c$ from the self-consistent Eqs.~(\ref{Eq:int_r_lock_real}) and~(\ref{Eq:int_r_drift_real}).

Following the same procedure for increasing coupling strength, we also sketch phase trajectories of two oscillators with degree $k=8$ and $20$, respectively, in the parameter space as shown in Fig.~\ref{fig:parameter_space_increasing_decreasing}(b). For high coupling strength, the population acts like a giant node and $r\simeq 1$. If $I<1$, only a stable fixed point exists, whereas the oscillators converge to fixed points. 
The oscillators with degrees $k \geq 20$ are dragged out of synchronization more easily. 
For $I>1$ the oscillators with degree $k = 20$ are easier to be out of synchronization compared to the ones with degree $k=8$. In this way, the order parameter $r$ would first slightly decrease  and then abruptly drop to lower values. 

\section{Analytical results and simulations}
\label{sec:analytical_results}
\begin{figure}[!htp]
\includegraphics[width=\linewidth]{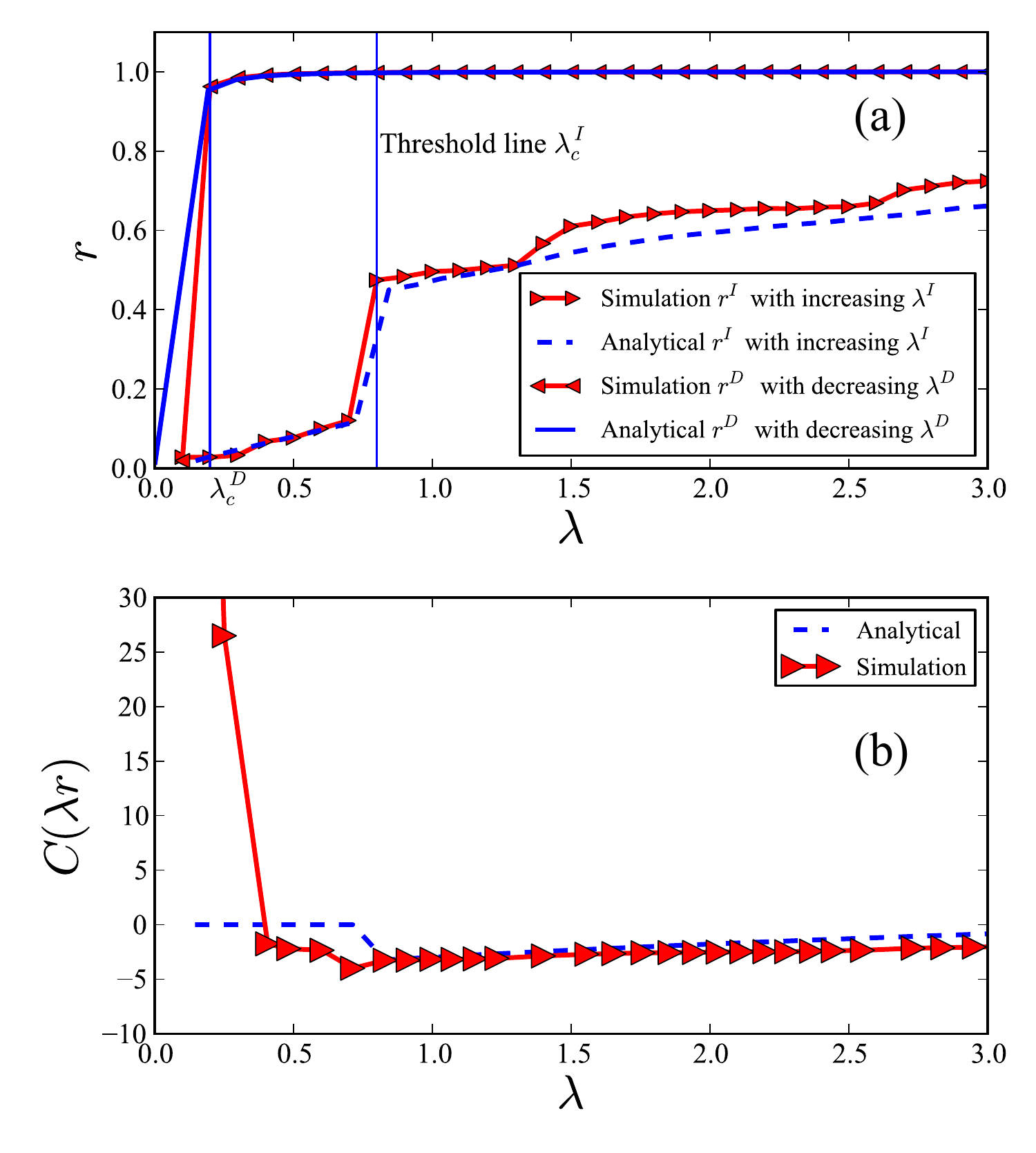}
\caption{(Color online). Analytical (in blue) and numerical (in red) analysis of the order parameter $r$ (a) and $C(\lambda r) $ with increasing coupling strength (b) for synchronization diagrams. We set the value $C(\lambda r)$ to be $0$ if $\lambda < \lambda_c^I$. The analytical plots are calculated from Eqs.~(\ref{Eq:r_lock_I}),~(\ref{Eq:int_r_drift_real}), and~(\ref{eq:equation_for_C}) with the synchronized boundary Eq.~(\ref{Eq:synchronized_boundary_increasing}) for increasing coupling and for decreasing coupling~(\ref{Eq:range_k_decrease}). Here the simulations are conducted with $\alpha=0.1$, $D=0.1$, and Barab\'{a}si-Albert scale-free networks characterized by
$N=3000$, $\left \langle k \right \rangle = 10$, and $k_{\min}=5$.}
\label{fig:order_parameter_coupling_strength}
\end{figure}
\subsection{Simulations on scale-free networks}
We demonstrate the validity of our mean-field analysis by conducting numerical simulations of the second-order Kuramoto model with $\alpha=0.1$ and $D=0.1$ on Barab\'{a}si-Albert scale-free networks characterized by
$N=3000$, $\left \langle k \right \rangle = 10$, $k_{\min}=5$ and the degree distribution $P(k) \sim k^{-\gamma}$, with $\gamma=3$.
Again, due to hysteresis, we have to distinguish two cases. First, we increase the coupling strength $\lambda$ from $\lambda_0$ by amounts of $\delta \lambda=0.1$, and compute the order parameter $r^I$ for $\lambda =
\lambda_0,\lambda_0 + \delta \lambda,...,\lambda_0+n\delta \lambda$.
Second, we gradually decrease $\lambda$ from $\lambda_0+n\delta \lambda$ to $\lambda_0$ in steps of $\delta \lambda$. Before each $\delta \lambda$-step, we integrate the system long enough ($10^5$ time steps) to arrive at stationary states, using a $4^{th}$ order Runge-Kutta method with time step $dt = 0.01$.

Figure \ref{fig:order_parameter_coupling_strength}(a) shows the synchronization diagrams for the model defined in Eq.~(\ref{Eq:Kuramoto_with_P}). The system exhibits the expected hysteretic synchrony depending on initial conditions. In the case of an increasing coupling strength $\lambda$, the initial drifting oscillators can be entertained to locked oscillators after certain transience. The order parameter $r$ remains at a low value until the onset of synchronization, $\lambda_c^I$, at which a first-order synchronization transition occurs, and $r$ increases continuously after that. In the case of decreasing $\lambda$, initially locked oscillators are desynchronized and fall into drift states once $\lambda$ crosses $\lambda_c^D$. For a high coupling strength, all oscillators are synchronized and $r=1$. As the coupling strength is decreased, the synchronized oscillators fall into unsynchronized states. As the two discontinuous transitions take place at different coupling thresholds, the order parameter exhibits hysteresis. 
\begin{figure}
\includegraphics[width=\linewidth]{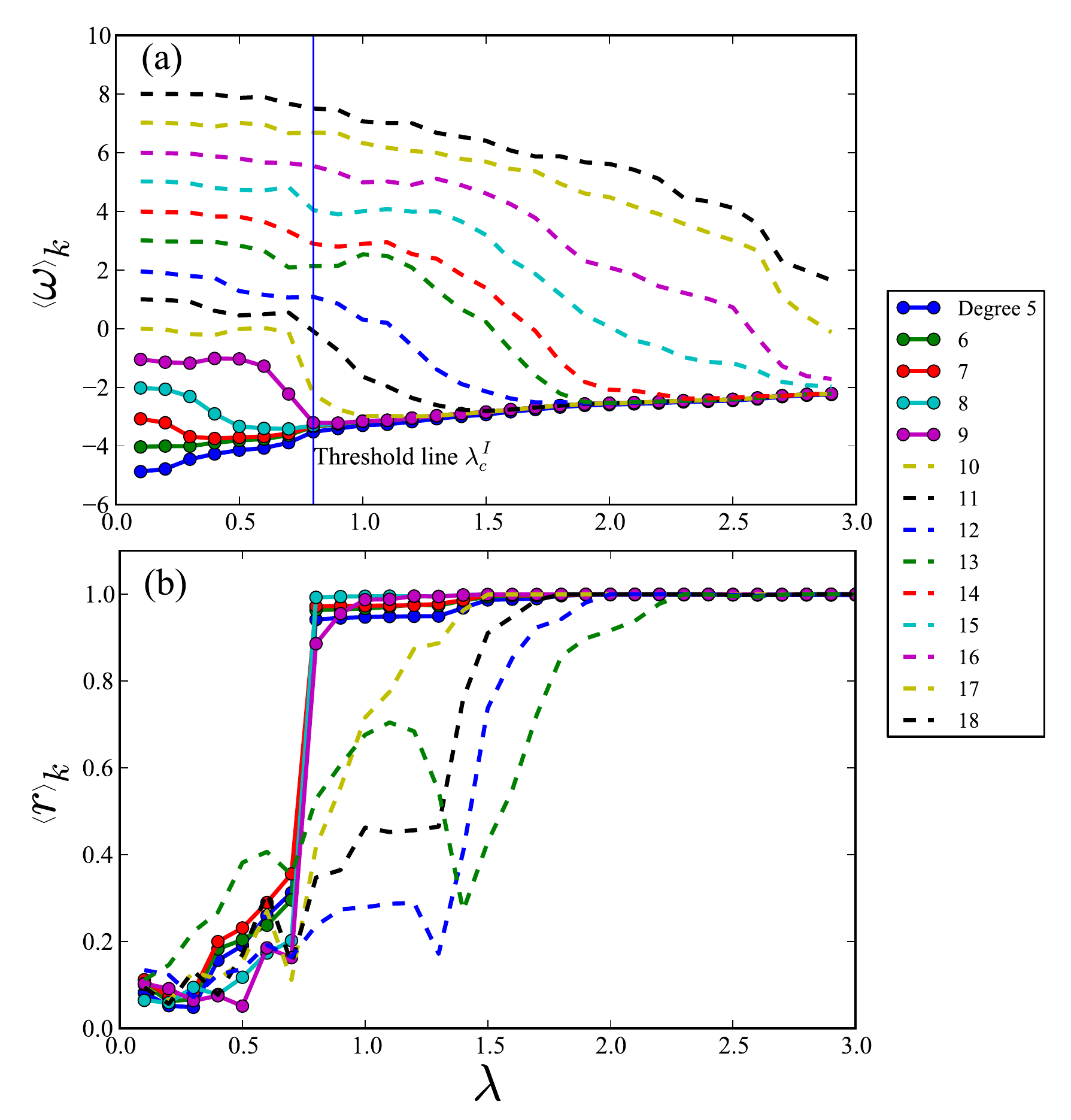}
\caption{(Color online). (a) Average frequency $\left\langle \omega \right\rangle_k$ and (b)  order parameter $\left\langle r \right\rangle_k$ of each cluster from simulations with $\left \langle k \right\rangle =10$. Solid lines denote synchronized clusters at the onset of synchronization. Dashed lines denote clusters composed of large degree nodes. The simulation parameters are the same as in Fig.~\ref{fig:order_parameter_coupling_strength}.}
\label{fig:frequency}
\end{figure}

To validate our mean-field analysis with simulation results, we simultaneously solve Eqs.~(\ref{Eq:r_lock_I}),~(\ref{Eq:int_r_drift_real}),~(\ref{eq:equation_for_C}), and (\ref{Eq:synchronized_boundary_increasing}) [(\ref{Eq:r_lock_I}),~(\ref{Eq:int_r_drift_real}),~(\ref{eq:equation_for_C}), and (\ref{Eq:range_k_decrease})] for increasing (decreasing) coupling strength. Note that the distribution of the natural frequencies is proportional to the degree distribution, and $\psi$ can not be set to a constant as has been done in previous works~\cite{strogatz2000kuramoto}. Recalling that $C(\lambda r)$ depends on $\dot{\psi}$ and $\ddot{\psi}$, we assume that $C(\lambda r) \approx 0$ when $\lambda < \lambda_c^I$, as each node oscillates at its own natural frequency. The oscillators with small degree synchronize first as shown in Fig.~\ref{fig:frequency}, and being in high percentage in a scale-free network, they dominate the mean field. The mean field rotates with a constant frequency $\dot{\psi}$. As before, it is convenient to analyze the system with $y\equiv \lambda r$ and $r=y/\lambda$. As we can see, the analytical results are in good agreement with the simulations.

To deepen the understanding of the transition to synchrony, we calculate the average frequency of all oscillators of degree $k$~\cite{PhysRevLett.110.218701}, $\left \langle \omega \right \rangle_k = \sum_{[i|k_i=k]} {\omega_i}/(NP(k))$, where 
$\omega_i=\int^{t+T}_t\dot\phi_i(\tau)dt / T$ and $t$ is large enough to let all oscillators reach stationary states. Figure~\ref{fig:frequency}(a) shows that each cluster, an ensemble of oscillators with same degree, oscillates independently before the onset of synchronization. Oscillators with small degree, denoted by a solid line, join the synchronous component simultaneously at $\lambda_c^I$. For further increasing coupling strength $\lambda$, more clusters, denoted by dashed lines, join the synchronized component successively according to their degrees, starting from smaller ones, and correspondingly $C(\lambda r)$ increases. 
\begin{figure}
\includegraphics[width=\linewidth]{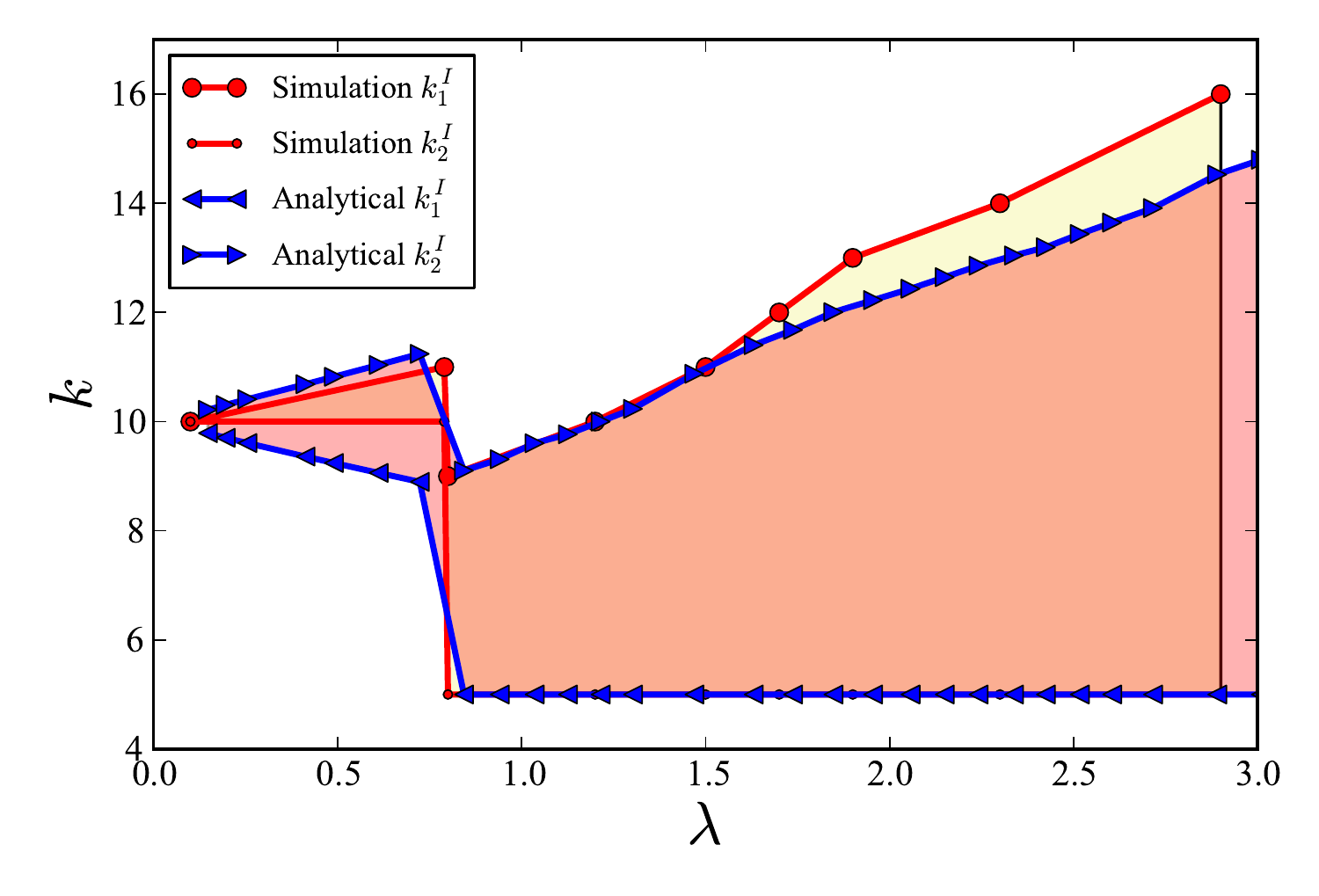}
\caption{(Color online). Synchronized degrees from analytical and simulation results with increasing coupling strength. The yellow (light gray) shading shows the range of synchronized degrees from the simulations and the red (dark gray) shading shows the range from the analytical results. }
\label{fig:synchronized_degree}
\end{figure}

What happens inside each cluster at the onset of synchronization? We define the order parameter of each cluster denoted by $\left\langle r \right\rangle_k$, $\left\langle r \right\rangle_k = \int_t^{t+T} r_k dt/T$, where $r_k e^{i\psi_k}=\sum_{[i|k_i=k]} e^{i \theta_k} /(NP(k))$. When $\lambda<\lambda_c^I$ and initial values of $\theta$ are selected at random from $[-\pi,\pi]$, the oscillators of each cluster follow the same dynamics. Therefore, the oscillators are uniformly distributed over the limit cycle and $\left\langle  r \right\rangle_k \approx 0$ as shown in Fig.~\ref{fig:frequency}(b). The order parameter of the synchronized clusters denoted by a solid line in Fig.~\ref{fig:frequency} jumps to $1$ at the onset of synchronization. After that, other clusters join the synchronized component and $\left\langle r \right\rangle_k$ approaches $1$ as denoted by the dashed lines.
\begin{figure}
\includegraphics[width=\linewidth]{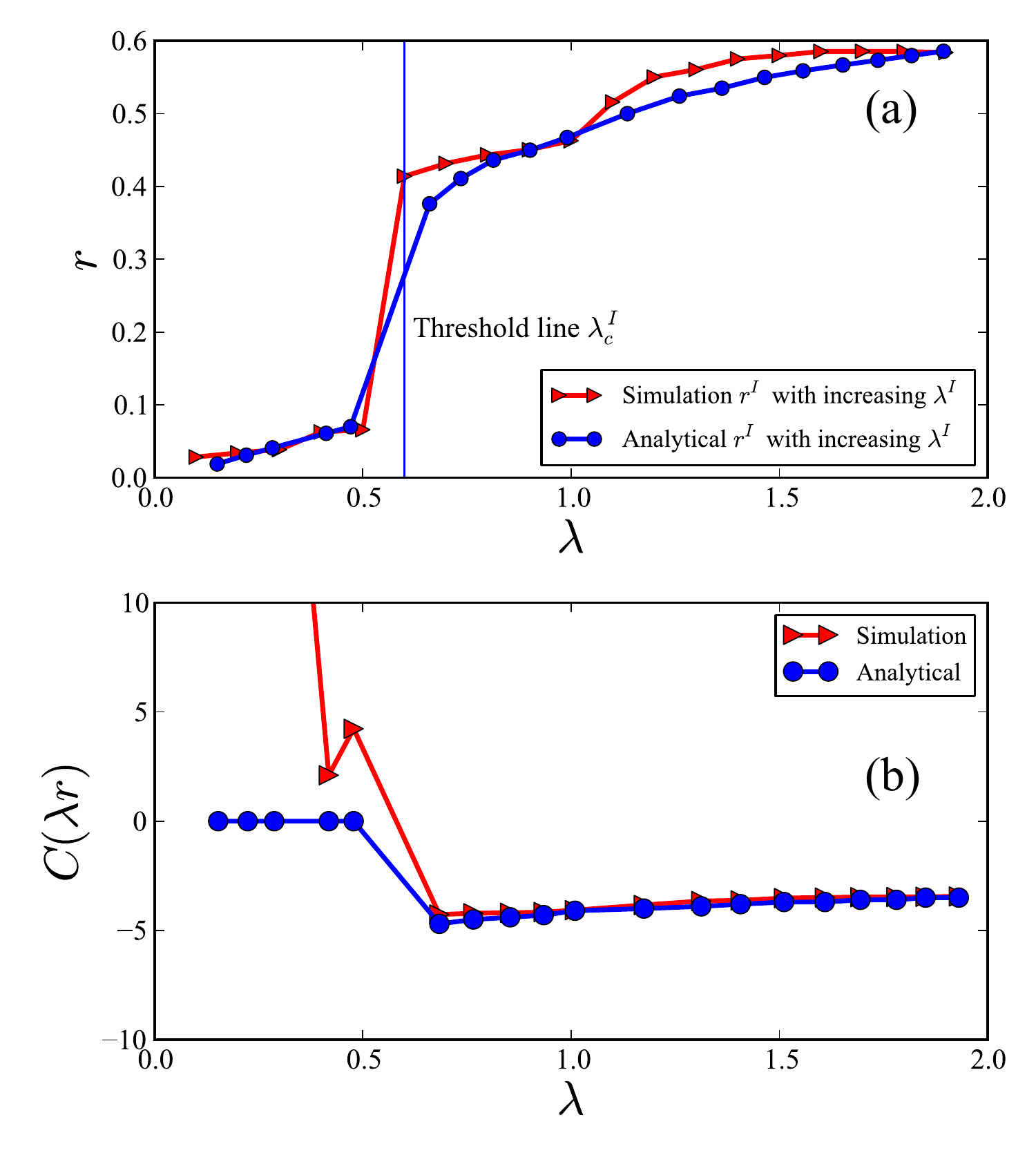}
\caption{(Color online). Results with increasing coupling strength $\lambda$. Part (a) shows the order parameter $r$ vs. $\lambda$. The red (blue) curve denotes the simulations (analytical) results. Part (b) shows the $C(\lambda r)$ vs. $\lambda$. The critical coupling is $0.6$. As in Fig.~\ref{fig:order_parameter_coupling_strength}(b), we take the value from solid lines. The analytical results are obtained from Eqs. (\ref{Eq:r_lock_I}), (\ref{Eq:int_r_drift_real}), (\ref{eq:equation_for_C}), and~(\ref{Eq:synchronized_boundary_increasing}). Here the simulations are conducted with $\alpha=0.1$, $D=0.1$ and Barab\'{a}si-Albert scale-free networks characterized by
$N=3000$, $\left \langle k \right \rangle = 12$, and $k_{\min}=6$. }
\label{fig:r_C_12}
\end{figure}

In Fig.~\ref{fig:synchronized_degree}, we show the synchronized boundary $k^I \in [k_1^I,k_2^I]$ as a function of the coupling strength $\lambda$ calculated from analytical expressions and extensive simulations for increasing $\lambda$. The analytical and simulation results are in good agreement. Note that the discontinuity of evolution of the synchronized boundary gives rise to a first-order phase transition in Fig.~\ref{fig:order_parameter_coupling_strength}(a). After the transition to synchrony, the low boundary $k_1^I$ stays constant at the minimal degree $k_{\min}=5$, and, as more clusters join the synchronized component, the upper boundary $k_2^I$ increases with $\lambda$. 

The above results are based on scale-free networks with the average degree $\left\langle k \right\rangle = 10$. 
To show more details, following the above process, we analyze the increasing coupling case with an average degree $\left\langle k \right\rangle = 12$ with minimum degree $k_{\min}=6$ as shown in Fig.~\ref{fig:r_C_12}. 
We integrate the equations~(\ref{Eq:r_lock_I}),~(\ref{Eq:int_r_drift_real}),~(\ref{eq:equation_for_C}) with~(\ref{Eq:synchronized_boundary_increasing}) and get the evolution of the $C(\lambda r)$ and the order parameter $r$ as a function of the coupling strength $\lambda$. We observe that the critical coupling strength in this case is smaller than that of scale-free networks with an average degree $\left\langle k\right\rangle=10$.

\begin{figure}
\includegraphics[width=\linewidth]{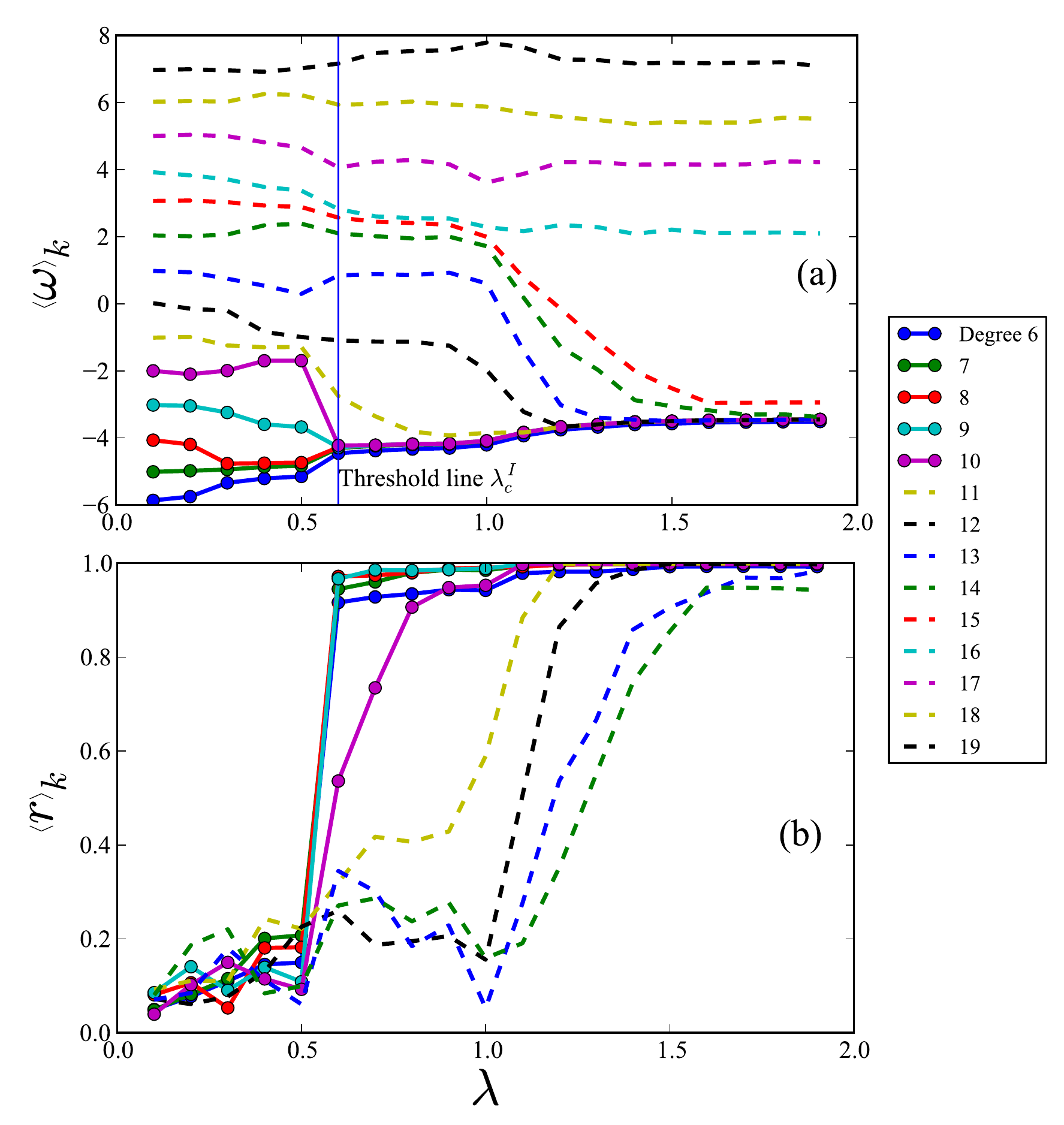}
\caption{(Color online). Results are obtained with the same parameter values as in Fig.~\ref{fig:r_C_12}. Part (a) shows the evolution of the average frequency of each cluster $\left\langle \omega\right\rangle_k$ as a function of $\lambda$, and (b) indicates the evolution of the order parameter of each cluster $\left\langle r \right\rangle_k$ of $\lambda$. Solid lines indicate the clusters synchronized at the critical threshold. The simulation parameters are the same as in Fig.~\ref{fig:r_C_12}. }
\label{fig:frequency_order_parameter_12}
\end{figure}

We follow the above process again and investigate the synchronization inside each cluster. 
As expected, initially oscillators for each cluster oscillate around its natural frequency and the order parameter $r$ for each cluster remains at a low value (Fig.~\ref{fig:frequency_order_parameter_12}). Increasing the coupling strength further, a first-order transition to synchronization occurs at the threshold $\lambda_c^I=0.6$. Clusters of nodes with a degree from $k=6$ to $k=10$ join the synchronization component simultaneously. More clusters join the synchronized component successively starting from low to high degrees.

\begin{figure}
\includegraphics[width=\linewidth]{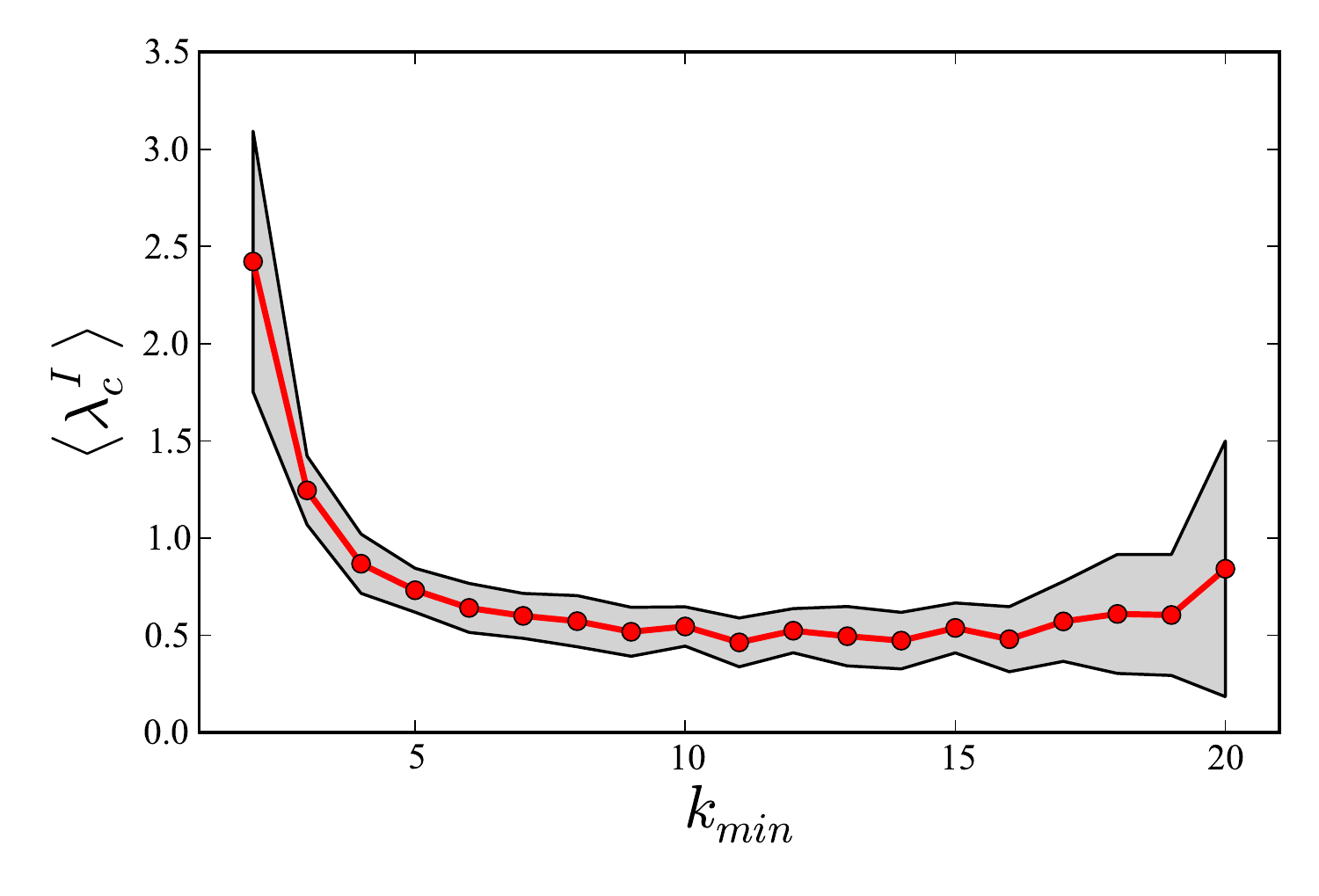}
\caption{(Color online). Mean values of critical coupling strength $\left\langle \lambda_c^I\right\rangle$ for increasing coupling with different minimal degrees $k_{\min}$. The gray shading indicates the standard deviation. Simulations at each minimal degree are conducted as in Fig.~\ref{fig:order_parameter_coupling_strength}. All networks have $N=3000$. }
\label{fig:mean_threshold_k_min}
\end{figure}

We also evaluate the influence of the average degree on the critical coupling threshold. Figure~\ref{fig:mean_threshold_k_min} shows the mean values of the critical coupling strength $\left\langle \lambda_c^I \right\rangle$ for increasing $\lambda$ with different minimal degrees $k_{\min}$ varying from $k_{\min}=2$ to $k_{\min}=20$. In simulations, we define a transition to synchrony if the difference between $r(\lambda)$ and $r(\lambda-\delta{\lambda})$ is larger than, for example, $0.1$. Due to the limitation of networks size, fluctuations of $\left\langle \lambda_c^I \right\rangle$ are unavoidable. The plots have been obtained with the same parameter values as above except minimal degrees $k_{\min}$. One can observe that the threshold values decrease with increasing minimal degrees initially and become almost constant afterwards. 

\begin{figure}[b!]
\includegraphics[width=1.0\linewidth]{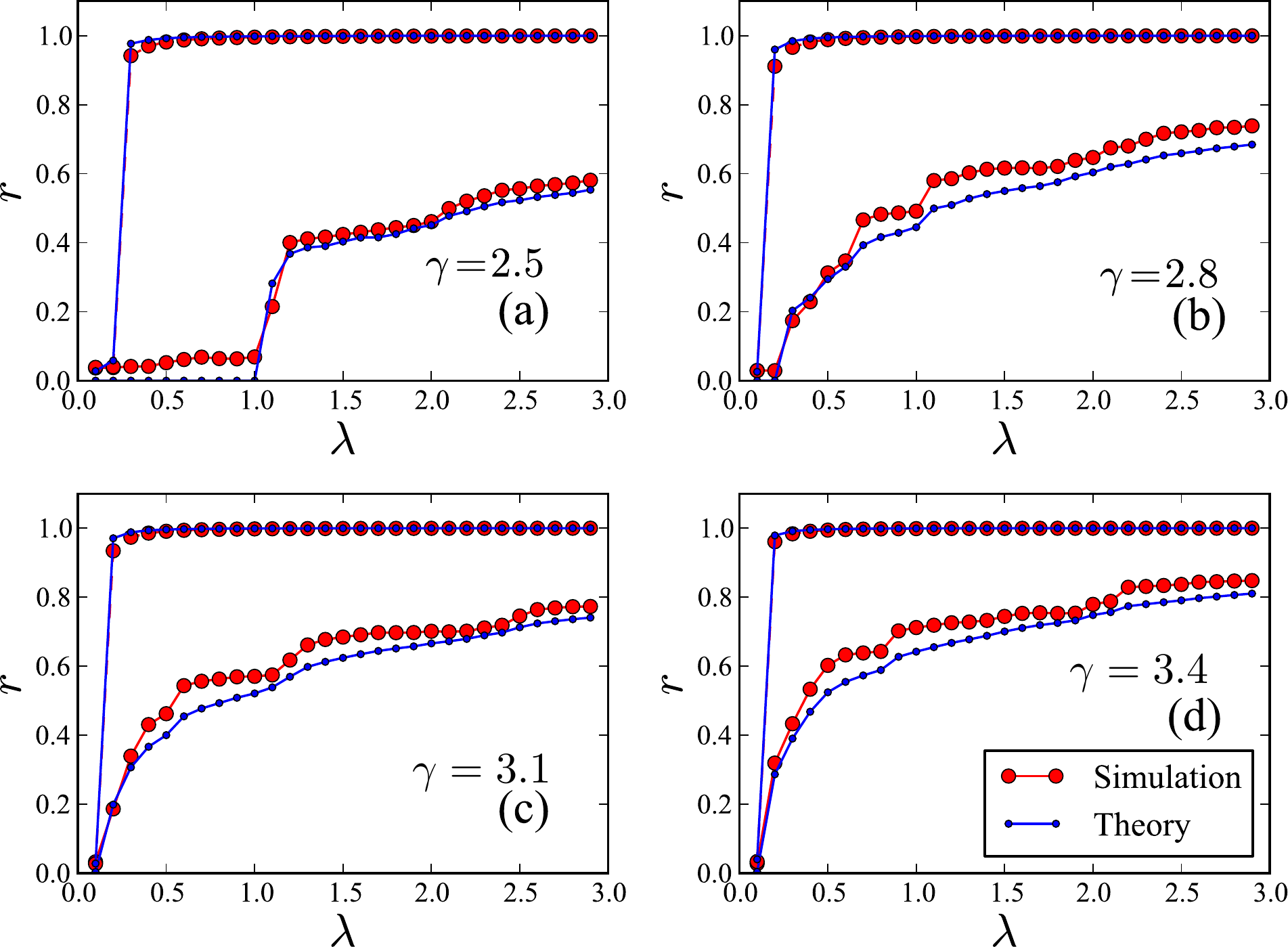}
\caption{ (Color online). Synchronization diagrams for networks with the degree-distribution $P(k) \sim  k^{-\gamma}$ for different exponents $\gamma$. The analytical plots are calculated from the summation of Eqs. (\ref{Eq:int_r_lock_real}) and (\ref{eq:r_drist_min_plus}). All networks have $N=3000$ and $\left\langle k \right\rangle = 10$.} 
\label{fig:configuration_model}
\end{figure}

To investigate the system's dynamical behavior in  networks with different levels of heterogeneity, in Fig.~\ref{fig:configuration_model} we present the synchronization diagrams for the forward and backward continuation of the coupling $\lambda$ for networks with 
degree distribution $P(k) \sim k^{-\gamma}$ considering different exponents $\gamma$. As expected, the 
onset of synchronization decreases for the forward propagation of the coupling strength $\lambda$, similarly to that observed in the first-order model~\cite{PhysRevLett.106.128701}. Interestingly, the branch associated to the backward propagation of the coupling $\lambda$ is barely affected by the changes of $\gamma$. A similar effect was recently reported in~\cite{olmi2014hysteretic}, where the authors observed a weak dependence of the critical coupling $\lambda_c^{D}$ on the network size $N$.

\subsection{Quenched disorder}

In the preceding section, we showed that abrupt transitions occur in scale-free networks of second-order Kuramoto oscillators, but the dependence of such discontinuous transitions on perturbations in the correlation between natural frequencies and topological properties is unknown. To address this question, here we consider the inclusion of quenched disorder on the natural frequencies in order to disturb such
correlations~\cite{zou2014basin,PhysRevE.89.062811}. More precisely, to check the robustness of cluster explosive synchronization, we set $\Omega_i = D(k_i - \left\langle k \right\rangle ) + \varepsilon_i$, where $\varepsilon \in [-q,q]$ is randomly drawn from a uniform distribution $g(\varepsilon)$. Therefore, the equations of motion in the continuum limit are given by
\begin{equation}
\ddot{\phi}=-\alpha\dot{\phi}+D\left[k-\left\langle k\right\rangle -C(\lambda r)\right]+\varepsilon-k\lambda r\sin\phi.
\label{eq:new_equations}
\end{equation}
As we increase the width of the distribution $g(\varepsilon)$, the topological influence on the natural frequency is decreased. 

We can calculate the contribution of locked oscillators $r^\textrm{q}_\text{lock}$ through
\begin{eqnarray}
r_{\textrm{lock}}^{\textrm{q}} & = & \frac{1}{\left\langle k\right\rangle } \int_{-q}^q \int_{k^\textrm{q}_{1}(\varepsilon)}^{k^\textrm{q}_{2}(\varepsilon)}kP(k)g(\varepsilon)\\
 &  & \times\sqrt{1-\left[\frac{D(k-\left\langle k\right\rangle -C(\lambda r))+\varepsilon}{k\lambda r}\right]^{2}}dk d\varepsilon, \nonumber
 \label{eq:r_lock_quenched}
\end{eqnarray}
where the degree range of the synchronous oscillators $k^{\textrm{q}}\in[k_{1}^{\textrm{q}}(\varepsilon),k_{2}^{\textrm{q}}(\varepsilon)]
$ in the presence of quenched disorder is determined by the conditions 

\begin{equation}
\frac{|D\left[k-\left\langle k\right\rangle -C(\lambda r)\right]+\varepsilon|}{k\lambda r}\leq 1,
\label{eq:quenched_condition_1}
\end{equation}
and
\begin{equation}
\frac{|D\left[k-\left\langle k\right\rangle -C(\lambda r)\right]+\varepsilon|}{k\lambda r}\leq\frac{4\alpha}{\sqrt{k\lambda r}}.
\label{eq:quenched_condition_2}
\end{equation}
Similarly, one can also get the contribution of drift oscillators denoted by $r^\textrm{q}_\text{drift}$ as follows

\begin{figure}
\includegraphics[width=\linewidth]{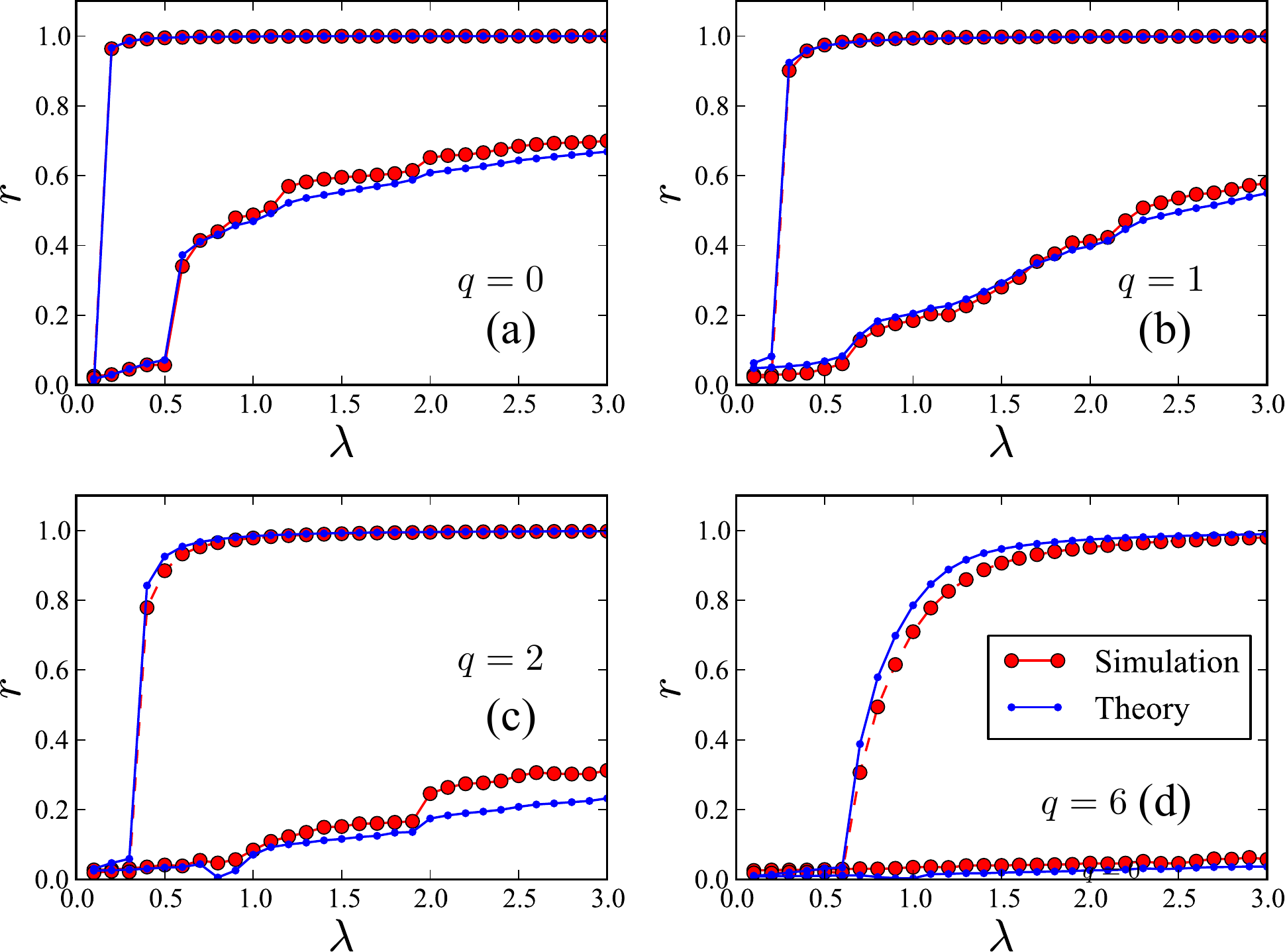}
\caption{(Color online). Synchronization diagrams with respect to different probabilities $q$ with $q=0$ (a), $q=1$ (b), $q=2$ (c), and $q=6$( d). Analytical results are obtained from the summation of $r^\textrm{q}_\text{lock}$ and $r^\textrm{q}_\text{drift}$.  Here, we use the same network topology as in Fig.~\ref{fig:order_parameter_coupling_strength} with $N=3000$, $\alpha=0.1$, $D=0.1$, and set $g(\varepsilon) = 1/2q$ with $\varepsilon \in [-q, q]$. }
\label{Fig:Quenched_disorder}
\end{figure}

\begin{widetext}
\begin{eqnarray*}
r_{\textrm{drift}}^{\textrm{q}} & = & -\int d\varepsilon\left(\int_{k_{\min}}^{k^\textrm{q}_{1}(\varepsilon)}dk+\int_{k^\textrm{q}_{2}(\varepsilon)}^{k_{\max}}dk\right)\frac{-rk^{2}\lambda\alpha^{4}P(k)g(\varepsilon) }{\left[D(k-\left\langle k\right\rangle -C(\lambda r))+\varepsilon\right]^{3}\left\langle k\right\rangle }\Theta\left(D\left[\left\langle k\right\rangle +C(\lambda r)-k\right]-\varepsilon\right)\\
 & + & \int d\varepsilon\left(\int_{k_{\min}}^{k^\textrm{q}_{1}(\varepsilon)}dk+\int_{k^\textrm{q}_{2}(\varepsilon)}^{k_{\max}}dk\right)\frac{-rk^{2}\lambda\alpha^{4}P(k)g(\varepsilon)}{\left[D(k-\left\langle k\right\rangle -C(\lambda r))+\varepsilon\right]^{3}\left\langle k\right\rangle }\Theta\left(D\left[k-\left\langle k\right\rangle -C(\lambda r)\right]+\varepsilon\right),
 \label{eq:r_drift_quenched}
\end{eqnarray*}
\end{widetext}
where $\Theta(\cdot)$ is the Heaviside function.
Figure~\ref{Fig:Quenched_disorder} shows the synchronization diagrams considering the same network configuration as in Fig.~\ref{fig:order_parameter_coupling_strength}, but taking into account different values for the quenched disorder. As we can see, the phase coherence of the lower branches decreases, enlarging the hysteresis area with increasing $q$. Interestingly, the upper branches decrease as the strength of the quenched disorder is increased and the onset of the transition increases accordingly. Therefore, the additional quenched disorder decreases the phase coherence and diminishes the abrupt transitions. 

\section{Conclusion}
\label{sec:conclusions}

In summary, we have shown that the cluster explosive synchronization happens in the second-order Kuramoto model presenting a correlation between natural frequency and degree, as verified for the first-order Kuramoto model~\cite{PhysRevLett.106.128701}. The synchronization diagram exhibits a strong hysteresis due to the different critical coupling strengths for increasing and decreasing coupling strength. As a function of the coupling strength, we have derived self-consistent equations for the order parameter. Furthermore, the projection of the phase transition on the parameter space of a pendulum has enabled the derivation of the analytical expression of the synchronized boundaries for increasing and decreasing coupling strength. 
We have solved the self-consistent equation and the synchronized boundaries simultaneously, and the analytical results have been compared to the simulations and both show a good agreement. 
 Moreover, following the same process, numerically and analytically, we have shown that the onset of synchronization for increasing coupling strength decreases with increasing scaling exponents but the onset of synchronization for decreasing coupling strength keeps constant.

To evaluate the robustness of abrupt transitions against the degree-correlated natural frequency, an additional quenched disorder is included. Numerically and analytically, we show that phase coherence and abrupt transitions decrease with the increasing of the strength of the quenched disorder.

The hysteresis in scale-free networks with different scaling exponents has also been investigated here, but the underlying mechanism for the occurrence of hysteresis in scale-free networks remains open. The impact of topology on dynamics with more sophisticated correlation patterns between local structure and natural frequencies as well as the formulation of the model considering networks of stochastic oscillators~\cite{sonnen2012onset} are subjects for further work.

\section*{Acknowledgements}

PJ would like to acknowledge China Scholarship Council (CSC) scholarship. 
TP would like to acknowledge FAPESP (No. 2012/22160-7) and IRTG 1740.
FAR acknowledge CNPq (grant 305940/2010-4) and FAPESP (grant 2011/50761-2 and 2013/26416-9) for financial support. 
JK would like to acknowledge IRTG 1740 (DFG and FAPESP) for the sponsorship provided.

\bibliographystyle{apsrev}
\bibliography{paper}
  
\end{document}